\documentclass{emulateapj} 
 
\usepackage[dvipdf]{epsfig} 
\usepackage[dvips]{rotating}
\usepackage{subfigure}
\usepackage{amsmath}
\usepackage{hyperref}

\bibpunct{(}{)}{;}{a}{}{,} 
 
\shorttitle{WISE Planet Nine Search}
 
\shortauthors{Meisner et al.} 
 
\begin{document} 
\title{Searching for Planet Nine with Coadded WISE and NEOWISE-Reactivation Images}

\author{Aaron M. Meisner\altaffilmark{1,2}}
\author{Benjamin C. Bromley\altaffilmark{3}}
\author{Peter E. Nugent\altaffilmark{2,4}}
\author{David J. Schlegel\altaffilmark{2}}
\author{Scott J. Kenyon\altaffilmark{5}}
\author{Edward F. Schlafly\altaffilmark{2,6}}
\author{Kyle S. Dawson\altaffilmark{3}}

\altaffiltext{1}{Berkeley Center for Cosmological Physics, Berkeley, CA 94720, 
USA}
\altaffiltext{2}{Lawrence Berkeley National Laboratory, Berkeley, CA, 94720, 
USA}
\altaffiltext{3}{Department of Physics \& Astronomy, University of Utah, Salt Lake City, UT, 84112, USA}
\altaffiltext{4}{Astronomy Department, University of California Berkeley, Berkeley, CA, 94720, USA}
\altaffiltext{5}{Smithsonian Astrophysical Observatory, 60 Garden Street, Cambridge, MA 02138, USA}
\altaffiltext{6}{Hubble Fellow}

\begin{abstract}
A distant, as yet unseen ninth planet has been invoked to explain various
observations of the outer solar system. While such a `Planet Nine', if it
exists, is most likely to be discovered via reflected light in the optical,
it may emit much more strongly at 3$-$5$\mu$m than simple blackbody
predictions would suggest, depending on its atmospheric properties 
\citep{Fortney16}. As a result, Planet Nine may be detectable at 3.4$\mu$m with
WISE, but single exposures are too shallow except at relatively small distances 
($d_9 \lesssim 430$ AU). We develop a method to search for Planet Nine far 
beyond the W1 single-exposure sensitivity, to distances as large as 800 AU, 
using inertial coadds of W1 exposures binned into $\sim$1 day intervals. We 
apply our methodology to $\sim$2000 square degrees of sky identified by 
\cite{Holman16} as a potentially likely Planet Nine location, based on the 
\cite{Fienga16} Cassini ranging analysis. We do not detect a plausible Planet 
Nine candidate, but are able to derive a detailed completeness curve, ruling
out its presence within the parameter space searched at $W1$$<$16.66 (90\% 
completeness). Our method uses all publicly available W1 imaging, spanning 2010
January to 2015 December, and will become more sensitive with future 
NEOWISE-Reactivation releases of additional W1 exposures. We anticipate that 
our method will be applicable to the entire high Galactic latitude sky, and we 
will extend our search to that full footprint in the near future.
\end{abstract}
 

\section{Introduction}

Beginning with the discovery of Sedna \citep{brown04}, distant trans-Neptunian 
objects (TNOs) with perihelia far beyond Neptune have posed a challenge to 
existing models of the solar system. Among other possibilities, \cite{brown04} 
suggested that Sedna's orbit may be indicative of an as yet undiscovered planet
in the outer solar system. Upon discovering the second sednoid, 2012 VP$_{113}$,
\cite{trujillo14} proposed the presence of a super-Earth perturber at
several hundred AU \citep[see also][]{sheppard16}. \cite{batygin16}
elaborated on this scenario by showing that a `Planet Nine' could simultaneously
account for sednoids and the clustering of distant TNOs, using
extensive simulations to significantly constrain the hypothetical planet's 
feasible mass range (5$-$20$M_{\oplus}$) and orbital parameters 
\citep[380 $<$ $a$/AU $<$ 980, ][]{brown16}.

As discussed in \cite{brown16}, Planet Nine, if it exists, is most likely
to be detected via reflected sunlight in the optical, where its
plausible range of brightnesses generally falls within reach of large
ground-based telescopes ($22 < V < 25$). Nevertheless, predictions
of detectability at other, longer wavelengths have also been made. 
\cite{Cowan16} showed that a Neptune-sized $\sim$40 K blackbody at many hundred
AU would be detectable at millimeter wavelengths with current and future 
CMB mapping experiments. \cite{linder16} suggested that Planet Nine would
likely be self-luminous. \cite{Fortney16} considered detailed atmospheric 
models, finding that for certain levels of methane depletion, Planet Nine could
have enormously enhanced 3$-$5$\mu$m emission relative to simple blackbody 
predictions. In the most optimistic models, Planet Nine would be detectable in
the WISE W1 \citep[$3.4\mu$m, ][]{wright10} channel, but would be too faint for
single exposure detection unless relatively nearby ($d_9 \lesssim 430$ AU).

Here we develop a method to search for Planet Nine using coadded W1
images, extending $\sim$1.4 magnitudes below the single-exposure detection 
limit. Although Planet Nine's infrared emission may well be orders of 
magnitudes too faint for detection by WISE, this full-sky dataset nevertheless
deserves to be systematically searched; W1 images spanning a nearly six year 
baseline have already been acquired and publicly released, and provide 
time sampling sufficient to enable Keplerian orbit linking. Moreover, the W1 
image quality is both extremely high in general, and very uniform with time 
\citep{neowiser, cutri15}. The WISE imaging data are therefore conducive to 
establishing fully-characterized constraints on Planet Nine's apparent infrared
brightness over most of the sky in the event of a nondetection.

Given the limited range of likely Planet Nine inclinations, its potential
paths across the sky are fairly well-constrained \citep{brown16}. Several
predictions have been made for Planet Nine's present-day location
\citep{Fienga16, Holman16, dlfm16}. \cite{Holman16} identified $\sim$1200
square degrees of southern sky toward the constellation Cetus as a 
high-probability Planet Nine location, by building on the Cassini ranging 
analysis of \cite{Fienga16}. Although alternative analyses of the Cassini 
ranging data remain ongoing, we nevertheless target the \cite{Holman16} 
preferred sky region, which at the very least provides a testbed area for 
developing and demonstrating our methodology.

The only WISE search for Planet Nine has been conducted by \cite{Fortney16},
filtering the AllWISE source catalog in/near the \cite{Holman16} favored 
sky region. However, the AllWISE catalog only utilizes WISE data spanning 
2010 January to 2011 February, representing just $\sim$1/3 of currently
available WISE imaging. Moreover, it was not possible to characterize the 
completeness of this search -- doing so would require simulating the AllWISE 
catalog-making software's interaction with a faint moving object that appears 
at certain epochs but disappears at others. 

Our basic search strategy is to generate inertial coadds of W1 exposures binned
into $\sim$1 day intervals, create a reference stack for each 
epochal coadd, extract a transient catalog via difference imaging, and finally 
link transients into Keplerian orbits. Searching for transients within 
time-resolved coadds allows for deeper and cleaner difference images than
would be possible with single exposures, though this choice does restrict the 
parameter space over which we are sensitive to TNOs. Because our search 
consists of a custom reprocessing that begins with raw WISE exposures, we can 
use all publicly available imaging, spanning 2010 January through 2015 
December, and can perform injection tests to thoroughly characterize our 
completeness.

In $\S$2 we briefly review the WISE and NEOWISE-Reactivation missions. In
$\S$3 we describe the design considerations and forecasted sensitivity of
our search method. In $\S$4 we list the data products that form the
basis for our search. In $\S$5 we detail our image coaddition
procedures. In $\S$6 we describe our difference imaging pipeline. In $\S$7 we
explain our orbit linking methodology. In $\S$8 we analyze our completeness.
In $\S$9 we discuss the results of our search, and conclude in $\S$10.

\section{WISE \& NEOWISE-Reactivation Overview}

The Wide-field Infrared Survey Explorer \cite[WISE; ][]{wright10} has surveyed
the full sky at four infrared wavelengths: W1=3.4$\mu$m, W2=4.6$\mu$m, 
W3=12$\mu$m and W4=22$\mu$m. The satellite resides in a $\sim$95 minute period 
Sun-synchronous low-Earth polar orbit, and the field of view is 
0.78$^{\circ}$$\times$0.78$^{\circ}$.

Launched in late 2009, WISE mapped the entire sky in all of W1-W4 between 7 
January 2010 and 6 August 2010.  As solid hydgrogen cryogen was exhausted, W4 
became unusable in early August 2010, as did W3 in late September 2010. During 
the NEOWISE mission phase \citep{neowise}, WISE continued surveying the sky in 
W1 and W2 until early February 2011, when the spacecraft was placed in 
hibernation. WISE was reactivated in October 2013, and returned to surveying 
the sky in W1 and W2. This W1/W2 survey, dubbed NEOWISE-Reactivation 
(NEOWISER), is planned to last through the end of 2016. NEOWISER first and 
second year data products, including all single-exposure images, have been 
publicly released in March 2015 and March 2016, respectively. NEOWISER images, 
especially in W1, have essentially the same sensitivity and overall high 
quality as those of the primary WISE mission \citep{neowiser}.

Because W1 single-exposure images represent the primary input to our search, 
we can make use of WISE data spanning all mission phases, from 2010 
January to 2015 December. Importantly, the W1 PSF is very broad relative to 
typical optical data, with 6.1$''$ FWHM. This means that 
Planet Nine could appear very nearly pointlike, even in W1 coadds which
bin exposures into $\gtrsim$1 day intervals.

\section{Survey Geometry/Cadence Considerations}
\subsection{WISE ``Visits''}
\label{sec:visits}
WISE surveys the sky by remaining pointed at very nearly 90$^{\circ}$ solar
elongation, completing one $360^{\circ}$ scan per orbit while approximately 
following a great circle of constant ecliptic longitude. As a result, away from
the ecliptic poles, a given sky region will intersect the WISE field of view at
six-monthly intervals, each time remaining in the $0.78^{\circ}$ FOV for 
$\sim$1 day, during which one exposure is acquired every $\sim$95 minutes. We 
refer to such day long periods of WISE observation as ``visits''.

Here we search for Planet Nine by creating per-visit inertial coadds of W1 
single exposures. Because we use data from all phases of the WISE+NEOWISER 
mission, we typically have 7 such epochal coadds available per sky location 
within our search footprint. This number of coadd epochs results from the fact 
that WISE has visited our search footprint during seven distinct ``seasons'', 
which we label with letters $a$-$g$ as defined in Table \ref{table:seasons}. A 
linkage of transients at four epochs along a Keplerian orbit is generally 
considered the minimum threshold to identify a TNO candidate. Therefore, our 
$\sim$7 coadded W1 epochs are sufficient to identify or place meaningful 
constraints on the presence of Planet Nine candidates. 

\begin{deluxetable}{ccclc}
\tablewidth{0pc}
\tablecaption{Season Definitions}
\tablehead{
\colhead{Season} & \colhead{MJD$_{min}$} & \colhead{MJD$_{max}$} & Calendar Months & \colhead{Parity}
}
\startdata
a & 55203 & 55239 & 2010 Jan $-$ 2010 Feb & 0 \\
b & 55370 & 55430 & 2010 Jun $-$ 2010 Aug & 1 \\
c & 55550 & 55593 & 2010 Dec $-$ 2011 Feb & 0 \\
d & 56648 & 56704 & 2013 Dec $-$ 2014 Feb & 0 \\
e & 56835 & 56895 & 2014 Jun $-$ 2014 Aug & 1 \\
f & 57012 & 57063 & 2014 Dec $-$ 2015 Feb & 0 \\
g & 57197 & 57256 & 2015 Jun $-$ 2015 Aug & \ 1
\enddata
\label{table:seasons}
\end{deluxetable}

Table \ref{table:seasons} also includes a ``parity'' column, which 
essentially serves as a binary indicator of Earth's location relative
to the Sun during each season. Because WISE always points at very nearly 
solar elongation of $90^{\circ}$, observations of a particular sky location
with the same parity have almost exactly zero relative parallax, whereas 
observations with opposite parity have relative parallax of approximately 2 AU.

The foregoing depiction of WISE's survey strategy is overly simplistic 
because it ignores Moon avoidance maneuvers, during which WISE can point many
degrees away from elongation 90$^{\circ}$. Our handling of 
these events is discussed in detail in $\S$\ref{sec:moon}. 
The remainder of this section's formulae and forecasts will ignore the
complication of Moon avoidance maneuvers, although in practice these actually
provide us with the benefit of extra coadd epochs. As a result,
we have on average 7.3 epochal coadds per sky location in the search region
studied.

\subsection{Parallax Within a Coadd Epoch}
\label{sec:parallax}
Apparent motion due to parallax during the course of a coadd epoch 
places restrictions on the range of distances at which a TNO will be 
detectable in our coadds, since moving sources become smeared out over an 
area larger than that of a point source, effectively decreasing their
signal-to-noise. To determine the extent of parallactic smearing, we
must first calculate the expected timespans of our epochal coadds.

Consider an idealized version of the WISE survey strategy, assuming WISE 
always points at exactly 90$^{\circ}$ solar elongation. At ecliptic
latitude (for which we use the symbol $\beta$) of zero, a given sky location 
will be observed over a period of 0.776$^{\circ}$$\times$(365.25 
days/360$^{\circ}$) = 0.787 days during a WISE visit. In reality, WISE does not 
point at precisely 90$^{\circ}$ elongation, but toggles back and forth by a few 
tenths of a degree on even/odd 
orbits\footnote{wise2.ipac.caltech.edu/docs/release/allsky/expsup/sec3\_4.html}. This effectively lengthens the $\beta$=0$^{\circ}$ visit duration to 
1.0$\pm$0.05 days, based on examination of per-pixel maps of minimum and 
maximum MJD for real coadds.

WISE obtains a constant number of frames per
unit ecliptic latitude, and therefore the number of exposures per unit
area within a visit scales as 1/cos($|\beta|$), as does the time spanned 
by exposures overlapping a given point on the sky:

\begin{equation}
\label{equ:timespan}
\tau_{coadd} = \frac{1.0 \ \mathrm{days}}{\mathrm{cos}(|\beta|)}
\end{equation}

Ignoring the curvature of the Earth's orbit, Planet Nine's parallax during
a single visit of WISE observations is:

\begin{equation}
\pi_9 = \frac{dl_{\perp}}{d_9} = \frac{(2\pi \ \mathrm{AU}) \times \frac{\tau_{coadd}}{365.25 \ \mathrm{days}} \times \mathrm{sin}(|\beta|)}{d_9}
\end{equation}

\noindent
Where $d_9$ is the distance to Planet Nine in AU, and the sin($|\beta|$)
factor projects the Earth's motion onto the direction perpendicular to
the line of sight. The combination of the previous two equations can be
rewritten as:

\begin{equation}
\label{equ:parallax}
\pi_{9} = 5.91'' \times \frac{600 \ \mathrm{AU}}{d_9} \times \mathrm{tan}(|\beta|)
\end{equation}

The smearing caused by parallactic motion means that there will be 
a \textit{minimum} distance at which our coadd-based search is sensitive
to Planet Nine. For instance, even in the absence of orbital drift, 
an object in the Kuiper belt at $|\beta|$=15$^{\circ}$ and 35 AU from the Earth 
would appear to move by $\sim$27$''$ within a coadd epoch, making
it very highly streaked. On the other hand at a fiducial
distance of 600 AU and the same $|\beta|$, Planet Nine would experience only 
1.6$''$ of parallactic motion, rendering its W1 profile very nearly pointlike.

\subsection{Effect of Parallactic Smearing}
\label{sec:neff}
To quantify and account for parallactic smearing, we compute the effective 
number of noise pixels of a TNO profile as a function of parallax within a 
coadd epoch. The effective number of noise pixels for a particular profile is 
given by:

\begin{equation}
n_{eff} = \frac{(\sum f_i)^2}{\sum f_i^2}
\end{equation}

\noindent
Where $f_i$ is the flux in each pixel $i$. We simulate Planet Nine profiles 
that translate by an amount $\pi_9$ during the course of a coadd epoch with 12
exposures, for 0$''$$<$$\pi_9$$<$8.25$''$. For these simulations, we use the W1
PSF constructed according to \cite{meisner14}, as shown in Figure 4  of 
\cite{lang14}. The resulting $n_{eff}$ values are well fit by:

\begin{equation}
\begin{split}
\frac{n_{eff}}{n_{eff}(\pi_9=0)} = 1 - (1.42\times10^{-3})\pi_9 \\ 
+ \ (9.59\times10^{-3})\pi_9^2 - (2.91\times10^{-4})\pi_9^{3}
\end{split}
\end{equation}

\noindent
With $\pi_9$ in arcseconds. 8.25$''$ corresponds to 3$\times$ the native WISE 
pixel size, and 1.35$\times$ the W1 FWHM, but even so 
$n_{eff}$($\pi_9$=8.25$''$) remains less than $1.5$$\times$ that of a W1 point 
source. We choose $\pi_9$=8.25$''$ as a threshold beyond which we will not 
claim any sensitivity of our search, since we do not aim to push into the 
regime of detecting/photometering highly elongated sources. We nevertheless 
believe that we have considerable sensitivity to less pointlike sources (see 
$\S$\ref{sec:latents}). We note that a typical galaxy 
($r_{half}$=0.45$''$) in 1$''$ seeing has $n_{eff}$ 1.99$\times$ larger than 
that of a Gaussian PSF -- by restricting to $\pi_{9}$$\le$8.25$''$, we remain 
strongly within the quasi-pointlike regime for which standard source detection 
and photometry techniques are applicable.

\subsection{Forecasted Sensitivity}
\label{forecast}

Our per-visit coadds very closely resemble the WISE Preliminary Release Atlas
stacks in terms of depth of coverage and sensitivity. In unconfused regions of 
the ecliptic plane with typical integer coverage (12 exposures), the 
Preliminary release 5$\sigma$ detection limit for static point sources is 
$W1_{lim,PSF}$ = 16.5 \citep{cutri11}. We always quote WISE magnitudes in the Vega system\footnote{wise2.ipac.caltech.edu/docs/release/allsky/expsup/sec4\_4h.html}. We can calculate $W1_{lim}$($|\beta|$, $d_9$) by scaling this static 
point source limiting magnitude to account for parallactic smearing via its 
influence on $n_{eff}$($|\beta|$, $d_9$), as well as the increasing number of 
exposures with increasing $|\beta|$:

\begin{equation}
\label{equ:depth}
\begin{split}
W1_{lim}(|\beta|, d_9) = W1_{lim,PSF} \ + \ \Delta_{ref} \\
 - \ 2.5\mathrm{log}_{10}\bigg[\sqrt{\frac{n_{eff}(|\beta|, d_9)}{n_{eff}(\pi_9=0)}} \sqrt{\frac{n_{exp}(\beta=0)}{n_{exp}(|\beta|)}} \bigg]
\end{split}
\end{equation}

\noindent
Where $n_{exp}$ is the number of W1 exposures contributing to a single sky 
location within an epochal coadd. Allowing $n_{exp}$ to represent an average 
capable of taking on fractional values, we adopt:

\begin{equation}
n_{exp}(|\beta|) = \frac{n_{exp}(\beta=0)}{\mathrm{cos}(|\beta|)}
\end{equation}

We will search for transients in difference images, rather than the epochal 
coadds themselves ($\S$\ref{sec:diffem}). The $\Delta_{ref}$ term represents a 
small loss in sensitivity to due to the noise added by subtracting the 
reference coadd, which typically has 6$\times$ more contributing exposures than
the epochal coadd. In that case:

\begin{equation}
\Delta_{ref} = -2.5\mathrm{log}_{10}\Big(\sqrt{(1 + 1/6)}\Big) = -0.084 \ \mathrm{mag}
\end{equation}

\begin{figure} 
 \begin{center}
  \epsfig{file=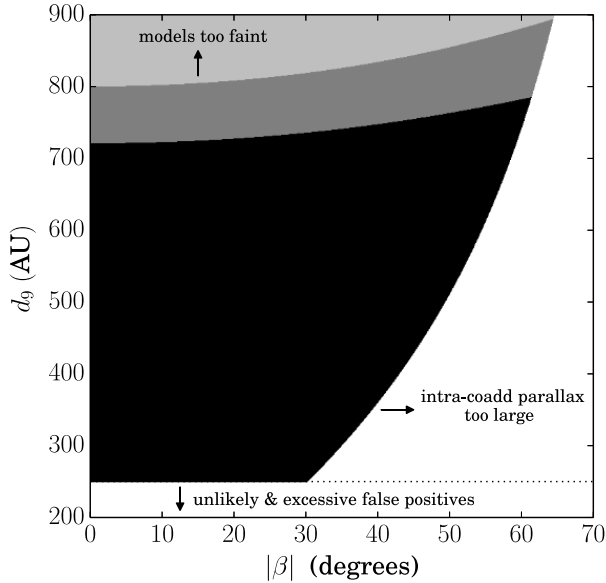, width=3.2in}
  \caption{\label{fig:mask} Parameter space accessible to our Planet Nine
search based on time-resolved W1 coadds. The black region is
expected to be accessible based on the forecast of $\S$\ref{forecast}. Both the
black and dark gray regions are accessible, based on the final
completeness analysis of $\S$\ref{sec:completeness}. The light gray area is
 also accessible, extending up to 2250 AU, but would require a
W1 luminosity brighter than the most optimistic \cite{Fortney16} model.}
 \end{center}
\end{figure}

We can now evaluate $W1_{lim}(|\beta|, \ d_9)$ and compare against 
Planet Nine's minimum W1 apparent magnitude as a function of $d_9$, yielding
a mask for accessible versus inaccessible regions of ($|\beta|$, $d_9$) 
parameter space. We calculate the minimum W1 magnitude as a function
of $d_9$ by assuming a maximum W1 luminosity of $H_{W1}$=2.13 \citep[$W1$=16.1 
at $d_9$=622 AU, ][Table 2]{Fortney16}. The results are shown in Figure 
\ref{fig:mask}.

Per the discussion in $\S$\ref{sec:neff}, regions with $\pi_9$$>$8.25$''$ are 
marked as inaccessible to our search in Figure \ref{fig:mask}. The $d_9$$<$250 
AU region of parameter space is also marked as inaccessible for multiple 
reasons. Planet Nine could only be so nearby if very close to perihelion, which
is unlikely \citep{brown16}. Additionally, the total apparent motion over 5.5 
years can be larger for smaller $d_9$, which requires fitting orbits to very 
large numbers of transient tuples. As a result, purely by chance, we expect to 
become disproportionately flooded with spurious linkages as we push toward 
lower $d_9$. Because the marginal benefit of pushing to $d_9$$<$250 AU is 
outweighed by the increase in false positives, we perform our search only for 
$d_9$$\ge$250 AU.

In summary, we expect substantial sensitivity to the maximally W1-luminous
\cite{Fortney16} model, especially at relatively low $|\beta|$. For instance, 
at $|\beta|$$\le$30$^{\circ}$ the $H_{W1}$=2.13 model is detectable at all 
distances between 250 and $\sim$700 AU. We also forecast sensitivity out to 
$d_9$$\approx$800 AU at high $|\beta|$. We note that during orbit linking 
($\S$\ref{sec:orbfit}), no upper limit is imposed on $d_9$. We are therefore
able to detect Planet Nine out to thousands of AU, provided it could be 
sufficiently more luminous than any of the \cite{Fortney16} models. Our actual 
sensitivity turns out to be $\sim$0.2 mag better than the present forecasts, 
essentially because of the distinction between the 5$\sigma$ detection 
threshold considered in this section and our 90\% completeness threshold for 
acquiring a quadruplet of detections ($\S$\ref{sec:completeness}). 

\begin{figure*}
 \begin{center}
  \epsfig{file=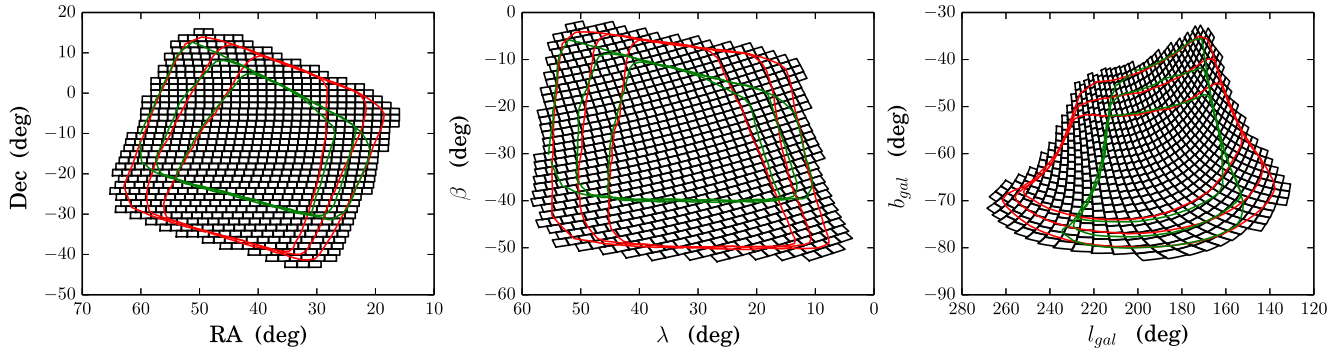, width=7.0in}
  \caption{\label{fig:footprint} Sky region searched in equatorial, ecliptic and Galactic coordinate systems. Black lines trace edges of the 805 unWISE tile
footprints which are analyzed throughout this work. Green and red contours
are those of \cite{Holman16}, Figure 9. These 805 unWISE tiles were chosen to 
fill the outer red contour, plus some margin. The area searched totals 1840 
square degrees.}
 \end{center}
\end{figure*}

Alternative search strategies based on linking single-exposure 
detections \citep[for instance a W1 variant of][]{luhman14} are not expected
to be sensitive beyond $d_9$=430 AU, given the single-exposure limiting
magnitude of $W1$=15.3 and adopting the maximum W1 luminosity from 
\cite{Fortney16}. Still, a search based on single-exposure transients would be 
complementary to our coadd-based analysis, filling in some of the low $d_9$
parameter space which is problematic for our approach.

\subsection{Drift within a Coadd Epoch}
\label{sec:drift}

We have thus far ignored the complication of orbital drift within the
timespan of a coadd epoch. If large enough, orbital drift could compromise our 
sensitivity by further smearing the appearance of Planet Nine in our coadds. To
determine the maximum intra-coadd orbital drift, we calculate the maximum 
intra-coadd drift within the mask shown in Figure 1, for each ($a_9$, $e_9$) 
pair in a set of orbits tracing the two loci of $a_9$ versus $e_9$ provided in 
$\S$2 of \cite{brown16}. We find that the maximum intra-coadd drift across all
orbits is $1.29''$, which occurs near $|\beta|$=$30^{\circ}$, for orbits 
currently very close to perihelion and with perihelia very similar to the 
inner edge of our search space ($q$ $\approx$ $d_9$ $\approx$ 250 AU).

Thus, the intra-coadd orbital drift is always expected to be
less than 1/4 of the W1 PSF FWHM. Because of the small amplitude of 
intra-coadd drift, and because it does not in general add constructively with 
parallactic motion, we do not attempt to account in detail for its minimal 
effect on our sensitivity.

\subsection{Drift Between Coadd Epochs}
\label{sec:d9_max}
We note that the upper limit of $d_9$ for which our $\S$\ref{sec:completeness} 
completeness curve calculation applies is dictated by \textit{lack} of 
sufficient orbital drift for very distant, eccentric objects near aphelion. 
Because a given coadd epoch's reference image ($\S$\ref{sec:ref}) will in
 general be constructed using other coadd epochs with zero relative parallax, 
an orbiting object drifting sufficiently slowly will behave somewhat like a 
static source, and at least partially cancel itself out in our difference 
images. Assuming $e\le0.75$, the annual drift\footnote{Annual drift is the 
relevant quantity because W1 coadd epochs with zero relative parallax are 
spaced almost exactly an integer number of years apart.} does not become 
smaller than the W1 FWHM until $d_9 \gtrsim 2250$ AU. Our pixel-level 
simulations suggest a negligible loss in sensitivity even in the case of 
6.1$''$/yr drift, and indeed the linear drift model of $\S$\ref{sec:ben} 
recovers faint high proper motion stars among our list of transients down to 
$\mu \approx  0.2''$/yr. To be conservative, we claim our completeness curve of 
$\S$\ref{sec:completeness} to be applicable only for $d_9 \le 2250$ AU, 
though we have considerable sensitivity at larger distances in the very 
unlikely event that Planet Nine could be so far away and detectably luminous.

\section{Input Data Details}
\label{sec:data}

The WISE ``Level 1b'' (L1b) single-exposure images constitute the primary input
data for our search. For every L1b frameset in the search region 
of interest, we employ the \verb|-w1-int-1b|, \verb|-w1-msk-1b| and 
\verb|-w1-unc-1b| images, which respectively provide the sky intensity 
and corresponding per-pixel bitmask and uncertainty values. We have 
downloaded a copy of these files for every publicly available frameset in 
our search region (shown in Figure \ref{fig:footprint}), including those from 
the primary WISE mission and the first two NEOWISER releases. In total, 
$\sim$315,000 W1 L1b framesets contribute to our analysis. The dates of 
observation of the WISE exposures analyzed range from 2010 January 7 to 2015
August 22 (UTC).

We also make use of the AllWISE Source Catalog and 2MASS All-sky Release 
source catalog \citep{skrutskie06}, to aid in masking static compact sources 
($\S$\ref{sec:filter}).

\section{Image Coaddition Methodology}
\label{sec:coadd}
To stack the W1 single exposures, we make use of the \cite{lang14}
unWISE coaddition framework. We employ the \cite{meisner16}
adaptation of the original unWISE codebase. The \cite{meisner16} unWISE 
coaddition pipeline includes updates to handle NEOWISER imaging and better 
correct time-dependent artifacts. We briefly review a few notable aspects of
the unWISE coaddition procedure here; refer to \cite{lang14} and 
\cite{meisner16} for a full discussion.

unWISE coaddition divides the sky into the same set of 18,240 
1.56$^{\circ}$$\times$1.56$^{\circ}$ astrometric footprints defined by 
the WISE team's Atlas stacks. The tile centers trace out a series of
iso-declination rings, and their axes are aligned along the equatorial cardinal 
directions. While the Atlas coadds are intentionally convolved by the WISE PSF,
our unWISE code preserves the native WISE angular resolution using Lanczos
interpolation, resulting in outputs with 2.75$''$ pixels and 2048 pixels on a 
side. We note that these footprints are not mutually exclusive. Neighboring 
tiles typically overlap each other by $\sim$3$'$ along each boundary.

In the \cite{lang14} and \cite{meisner16} coadds, there is no notion of
a coadded epoch like the one we introduced in $\S$\ref{sec:visits}. In 
these previous works, every W1 coadd is simply identified by its 
\verb|coadd_id| value, which is a string encoding the tile
center RA and Dec (e.g. `0000p000'). For our epochal coadds, we define a 
zero-indexed integer called \verb|epoch| such that a W1 epochal coadd is 
uniquely defined by its (\verb|coadd_id|, \verb|epoch|) pair. For each 
\verb|coadd_id|, \verb|epoch|
increases with time. However, a particular \verb|epoch| value will not in 
general correspond to a single season for all \verb|coadd_id| values; 
\verb|epoch|=1 of a given \verb|coadd_id| may occur in a different season than
\verb|epoch|=1 of another \verb|coadd_id|.

To create the time-resolved coadds analyzed in this work, 
as opposed to the full-depth stacks of \cite{meisner16}, we have made the 
following additional updates to the unWISE pipeline:

\begin{enumerate}
\item We segment the exposures which overlap each \verb|coadd_id| footprint
by MJD, into a series of $\sim$1 day coadd epochs. The boundaries of each
coadd epoch coincide by default with the beginning and end of a WISE visit.
However, a visit affected by a Moon avoidance maneuver is split
into two when enforcing the rules stipulated in $\S$\ref{sec:moon}.
\item Each quadrant of \textit{every} L1b exposure receives a fourth-order 
polynomial background correction as described in $\S$6 of \cite{meisner16} 
before being accumulated into its epochal coadd. The \cite{meisner16} 
full-depth W1 stacks are used as templates with respect to which the 
polynomials are computed. These corrections help to avoid sharp edge-like 
features in our coadds near the boundaries of contributing L1b exposures, which 
could potentially lead to large numbers of spurious difference image detections.
\item After creating an initial version of each epochal coadd, we
generate a final version by ignoring any exposures which were flagged as
having an excessive fraction of outlier pixels ($>$1\%) during creation of the 
initial coadd. Such exposures are rejected midway through the first iteration 
of coaddition, but their outliers compromise our flagging of outlier 
pixels in \textit{other} exposures. By performing a second coadd iteration that 
completely ignores problematic exposures from the start, we are able to
greatly reduce the number of spurious transients due to cosmic rays, 
particularly in sky regions affected by the South Atlantic Anomaly.
\item We generate two additional metadata images per epochal coadd. First,
we create a map of the mean MJD of the contributing exposures at each pixel.
This data product is crucially important, because its values at the
positions of detected transients are used during Keplerian orbit linking
($\S$\ref{sec:orbfit}). Second, we make a map of the time spanned by the 
exposures contributing to each coadd pixel. We use these maps to ensure that 
the coadd pixel timespans never significantly exceed the expectation from 
Equation \ref{equ:timespan}.
\end{enumerate}

\begin{figure}
 \begin{center}
  \epsfig{file=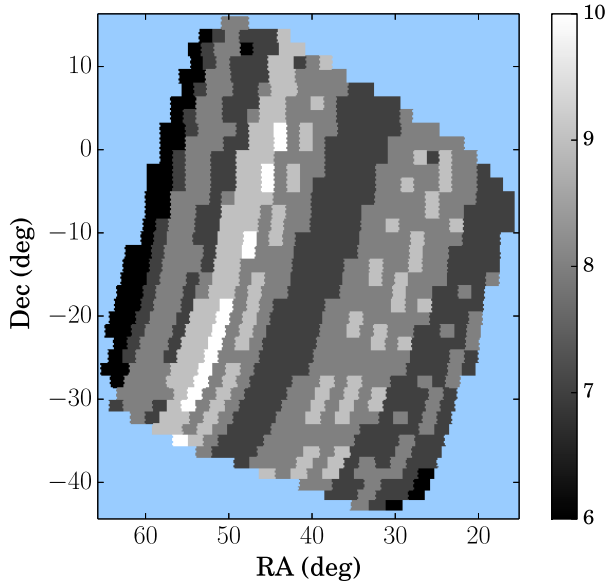, width=3.2in}
  \caption{\label{fig:num_epochs} Map of the number of coadd epochs 
per \texttt{coadd\_id} footprint throughout our search region.}
 \end{center}
\end{figure}

The search region shown in Figure \ref{fig:footprint} consists of 805 distinct 
\verb|coadd_id| astrometric footprints. In total, we generated 6,219 epochal 
coadds. Each \verb|coadd_id| footprint has between 6 and 10 epochs available. 
The number of coadd epochs per \verb|coadd_id| footprint throughout our search 
region is shown in Figure \ref{fig:num_epochs}.

\begin{figure}
 \begin{center}
  \epsfig{file=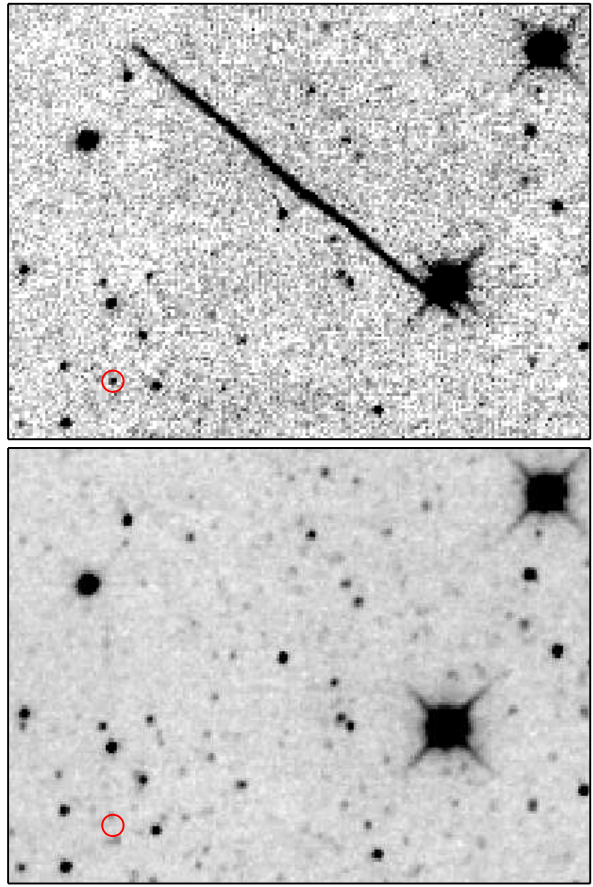, width=3.2in}
  \caption{\label{fig:l1b_vs_coadd} Small 9.0$'$$\times$6.7$'$ 
field extracted from epochal coadd with (\texttt{coadd\_id}, \texttt{epoch}) = 
(0211p000, 1) and one of its contributing exposures, illustrating the benefits 
of our coaddition process. Top: contributing W1 exposure 06337b163. Bottom: 
time-resolved coadd resampled onto the 06337b163 astrometry. Both panels have 
identical grayscale stretches. The coadd is visibly much deeper and less noisy,
as expected. In addition, a satellite streak which would likely lead to many 
spurious transient detections in the L1b image has been completely eliminated. 
The coaddition is also highly effective at rejecting fast-moving main-belt 
asteroids -- the red circle marks the location of (6136) Gryphon during 
exposure 06337b163.}
 \end{center}
\end{figure}

Figure \ref{fig:l1b_vs_coadd} shows an example of the benefits of our
coadds relative to single exposures. The coadds are $\sim$1.2 magnitudes
deeper than L1b images, and the numerous main-belt asteroids and
satellites found in single exposures are completely removed by the 
unWISE outlier rejection.

\subsection{Outlier Rejection Tests}
One might reasonably ask whether the same outlier rejection that 
removes satellite streaks and main-belt asteroids could also erase Planet Nine
from our epochal coadds. We have performed injection tests to 
address this concern. We consider the worst-case (fastest-moving) 
scenario by injecting point sources with parallax of 8.25$''$/day into 
individual exposures, and then running the modified exposures
through our coaddition process. We inject 100 such fakes per W1 
magnitude, for dozens of W1 magnitudes spanning 14.0$\le$$W1$$\le$16.8. The
outlier rejection should be most problematic for the brightest Planet Nine
injections. This motivated our choice to extend these fakes as bright as
$W1$=14.0, since the maximally W1-luminous \cite{Fortney16} model
would have $W1$$\approx$14.2 at the inner edge of our search space 
($d_9$=250 AU).

We find that the outlier rejection causes a very small, but nonetheless
measurable, reduction in the average fluxes we recover relative
to the true fluxes of the injections. Near our 90\% completeness threshold of 
$W1$=16.66 ($\S$\ref{sec:completeness}), on average 7 mmag of flux is lost. The 
fraction of flux lost ramps up slowly for brighter injections. At the bright 
end, $W1$=14.0, 19 mmag of flux is lost on average.

Because the impact of unWISE outlier rejection on a maximally fast-moving 
Planet Nine near our 90\% completeness threshold is less than a hundredth
of a magnitude, we neglect this effect in our completeness analysis. In short, 
Planet Nine is expected to be too slow-moving and faint to trigger unWISE 
outlier rejection at a level which significantly impacts our sensitivity.

\subsection{Moon Avoidance Maneuvers}
\label{sec:moon}
The largest departures of WISE's pointing relative to solar elongation of 
90$^{\circ}$ occur during so-called ``Moon-avoidance maneuvers''. These
maneuvers coincide with the roughly monthly occasions on which the Moon 
crosses the WISE 90$^{\circ}$ elongation sight line. At the start of
such a maneuver, WISE will gradually shift several degrees ahead in ecliptic 
longitude ($\lambda$) relative to 90$^{\circ}$ elongation, then avoid the Moon 
by jumping $\sim$5-10$^{\circ}$ backward in $\lambda$. This backward jump in 
ecliptic longitude has the effect of lengthening WISE visits near in time to a 
Moon avoidance maneuver to last up to $\sim$5-7 days. Coadds spanning such 
long periods of time would dramatically dilute the signal from an object 
orbiting at hundreds of AU, which could undergo many dozens of arcseconds of 
parallax during such a timespan.

In order to avoid long coadd timespans that are strongly inconsistent
with Equation \ref{equ:timespan}, we enforce the following rules:

\begin{enumerate}
\item No epochal coadd may include exposures during a Moon avoidance maneuver. 
A list of start/end MJD values for the maneuvers relevant to our search region
is provided in Table \ref{table:moon}. By discarding the frames acquired during
Moon avoidance maneuvers, we lose 1.6\% of exposures.

\item No epochal coadd may include exposures acquired both before and after the
same Moon avoidance maneuver.

\end{enumerate}

This time-slicing splits the 5-7 day visits created by Moon avoidance 
maneuvers into two short coadd epochs which span lengths of time consistent 
with or shorter than Equation \ref{equ:timespan}. As a result, even though 
there are only 7 seasons of W1 observations available, we end up with an 
average of 7.7 coadd epochs per \verb|coadd_id|, thanks to the ``extra'' 
epochs created by Moon avoidance maneuvers. Once the above rules are 
enforced, the MJD ranges spanned by exposures contributing to actual pixels in 
our epochal coadds closely follow Equation \ref{equ:timespan}. The largest 
measured value of $\tau_{coadd}$ for any pixel in our analysis is 1.65 days,
which occurs at $\beta$=$-52.3^{\circ}$ (the expectation
from Equation 1 at this $|\beta|$ is 1.64 days). The mean measured value of 
$\tau_{coadd}$ across all pixels in our analysis is 1.01 days.

Note that WISE Moon avoidance maneuvers are spaced such that a single 
\verb|coadd_id| can have at most two coadd epochs during a single season. For 
cases in which one \verb|coadd_id| has two coadd epochs in one season, these 
epochs are separated by a small handful of days.

\begin{deluxetable}{cllc}
\tablewidth{0pc}
\tablecaption{Moon Avoidance Maneuvers}
\tablehead{
\colhead{Season} & \colhead{MJD$_{start}$} & \colhead{MJD$_{end}$} & \colhead{$\lambda_{approx}$}
}
\startdata
a & 55219.5  & 55220.075 & 38.0$^{\circ}$ \\
b & 55381.28 & 55381.88 & 12.0$^{\circ}$ \\
b & 55410.91 & 55411.53 & 41.0$^{\circ}$ \\
c & 55573.63 & 55574.18 & 26.5$^{\circ}$ \\
d & 56665.1 & 56665.62 & 20.0$^{\circ}$ \\
d & 56694.815 & 56695.27 & 50.0$^{\circ}$ \\
e & 56856.71 & 56857.16 & 23.0$^{\circ}$ \\
e & 56886.10 & 56886.55 & 51.0$^{\circ}$ \\
f & 57019.74 & 57020.21 & 9.5$^{\circ}$ \\
f & 57049.49 & 57049.97 & 43.0$^{\circ}$ \\
g & 57211.67 & 57212.08 & 16.0$^{\circ}$ \\
g & 57240.87 & 57241.335 & \ 43.5$^{\circ}$
\enddata
\label{table:moon}
\end{deluxetable}

\subsection{Creating Reference Stacks}
\label{sec:ref}
For each time-resolved coadd identified by its unique
(\verb|coadd_id|, \verb|epoch|) pair, we generate a corresponding reference 
stack. To do so, we combine all epochal coadds for the relevant 
\verb|coadd_id| astrometric footprint, excluding the epoch under 
consideration. We also exclude any other epoch during the same season as the 
epoch for which the reference is being created. This latter constraint is
important because, in cases where a single \verb|coadd_id| has two epochs 
during the same season, these epochs are typically separated by only a couple 
of days, and Planet Nine's apparent motion could be slow enough that its 
detections at these two epochs have some overlap. To combine the epochal coadds
into a reference image, we apply a weighted median filter on a per-pixel basis.
The mean integer coverage of our reference stacks is 71.5 exposures, 
$\sim$6$\times$ the typical integer coverage of our epochal coadds.

\section{Difference Imaging Pipeline}
\label{sec:diffem}

We use a suite of standard image differencing and source extraction software to 
obtain a catalog of transient detections in each epochal coadd.

\subsection{Generating Difference Images}
We begin by astrometrically calibrating the reference coadd to 2MASS
using SCAMP \citep{scamp}. We next use SCAMP to astrometrically calibrate
the epochal coadd to 2MASS by comparing its source positions against the 
2MASS-calibrated reference stack source positions. We then use
SWARP \citep{terapix} to warp the reference and its uncertainty image to match
the astrometry of the epochal coadd.

We use the warped reference to create two difference images for each epochal
coadd. First, we perform a subtraction using the HOTPANTS 
package\footnote{www.astro.washington.edu/users/becker/v2.0/hotpants.html}, which applies the \cite{alard00} kernel-matching technique. Second, we 
perform a direct subtraction without kernel matching. We find that for WISE,
which delivers highly stable space-based imaging, kernel matching does
not play the same crucial role it does for ground based images,
where the PSF shape/size can vary dramatically. When creating the direct
subtraction, we do take into account small zero point differences
between the epochal and remapped reference images, using the \verb|KSUM00| value
obtained by HOTPANTS.

\subsection{Extracting Transient Sources}
To extract transient sources from the subtracted images, we use 
Source Extractor \citep{sextractor} in ``dual image mode'', which
makes use of static source properties in the remapped reference image to 
inform source detection in the subtracted image.

\subsection{Filtering Difference Image Extractions}
\label{sec:filter}
In addition to legitimate transients, the source extraction performed on
our difference images also yields large numbers of false positives. 
We need to cull the list of transients for several reasons. First, the
number of quadruplets that must be fit with Keplerian orbits during
linking scales as the fourth power of transient number density, meaning
that our full list of difference detections would be computationally
expensive to link. Aside from computational limitations, we also expect
the number of chance occasions on which spurious transients happen to link
into Keplerian quadruplets to scale as the fourth power of the transient number
density. Thus, eliminating as many spurious transients as possible from the 
outset allows us to obtain a manageable number of candidate quadruplets 
well-fit by Keplerian orbits.

To remove false positives, we would ideally employ
an approach like that of \cite{goldstein15}, analyzing a variety of
pixel-level features for each transient in order to classify it as legitimate
versus spurious. Unfortunately, that methodology has thus far only been 
developed for the case of pointlike transients, whereas we are explicitly 
searching for a moving object which may appear slightly streaked. We
therefore resort to a number of heuristics. These heuristics are designed
to have minimal impact on our completeness, and have effects which can
be easily incorporated into our simulations of $\S$\ref{sec:completeness}.
The following subsections detail each of the transient filtering heuristics.

\subsubsection{2MASS Sources}
\label{sec:tmass}
We discard any transient within a 1 pixel radius (2.75$''$) of a 2MASS
source location. This cut eliminates 0.16\% of the search area.

\subsubsection{Bright Static W1 Sources}
\label{sec:bright}
We create a mask for each bright W1 compact source in the AllWISE catalog with 
\verb|w1mpro|$<$9.5, and discard all transients that land within these masks.
The mask for each bright source is built by determining which pixels in the
coadd will be brighter than a threshold value of 100 Vega 
nanomaggies\footnote{A nanomaggie is a linear unit of flux such that
a 1 nanomaggie source corresponds to a magnitude of 22.5.}, given the source's 
total flux, centroid and the \cite{meisner14} W1 PSF model. These masks include
considerable detail, as the 14.9$'$$\times$14.9$'$ \cite{meisner14} PSF model 
extends quite far into the wings. For instance, our bright star masks capture 
significant portions of the diffraction spikes. These masks eliminate 0.46\% of
the search area.

\subsubsection{Kernel Matching Artifacts}
\label{sec:direct}
We find that the HOTPANTS kernel matching produces ``ringing'' features
around moderately bright sources, and that these artifacts are frequently 
identified as difference image detections. These are difficult to mask 
without discarding an excessive amount of area because they often extend 
$\sim$10-15$''$ from their parent source's centroid. We find that when running 
source detection on difference images created via direct subtraction, virtually
all legitimate transients are still recovered, while the spurious detections 
associated with moderately bright stars are strongly confined near their 
centroids. Therefore, we require that transients be detected in both the 
HOTPANTS and direct subtractions (using a 5.5$''$ match radius). This 
requirement does reduce the overall depth of our search, as the noise 
properties of the difference images with and without kernel matching can be 
different. Using the simulations described in $\S$\ref{sec:completeness}, we 
find the size of this effect to be 0.048 mags. On the other hand, this 
requirement leads to a $\sim$2.5$\times$ reduction in the number of transients,
leading us to deem the depth trade-off acceptable. Note that, although we 
create two separate difference detection catalogs per epochal coadd, our
downstream analyses always employ transient properties (in particular RA and 
Dec) measured from HOTPANTS subtractions with kernel matching.

\subsubsection{Reference Image Brightness}
\label{sec:refcut}
\begin{figure} 
 \begin{center}
  \epsfig{file=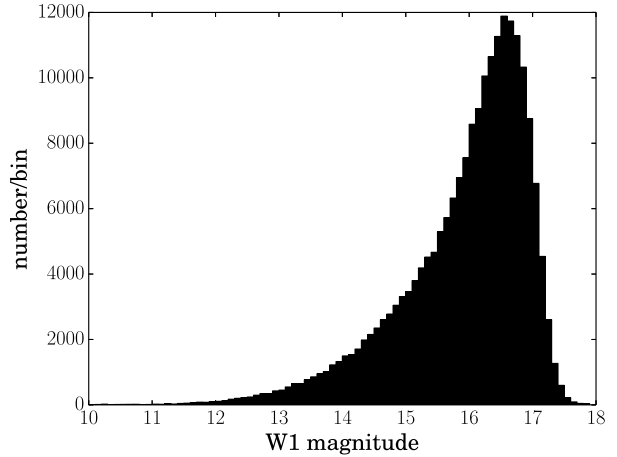, width=3.2in}
  \caption{\label{fig:mag_hist} Histogram of W1 magnitudes for detections in
our filtered transient catalog. The source counts peak near $W1$=16.6.}
 \end{center}
\end{figure}

We reject any transient with \textit{total} flux less than
the reference coadd brightness at its centroid. Given the W1 PSF and our
coadd pixel size, this is roughly equivalent to rejecting any source with 
peak brightness fainter than the reference image at the same location by
a factor of more than 7. For the transient flux, we adopt the Source Extractor 
\verb|flux_best| measurement.  It is not entirely straightforward to translate 
our reference brightness cut into a fraction of area masked, but we note that 
the transient magnitude histogram (Figure \ref{fig:mag_hist}) is sharply peaked
near $W1$=16.6, and for this magnitude 0.48\% of the total search area is 
masked.

\subsubsection{Negative Fluxes}
We discard transient detections with negative \verb|flux_best|, as these are 
generally spurious extractions near bright source residuals.

\subsubsection{Ultra-Bright Stars}
\label{sec:ultrabright}
A small handful of stars are so exceptionally bright that the approach of 
$\S$\ref{sec:bright} is insufficient, and such stars result in localized
regions with extremely high transient number density. In these cases, we 
construct special masks consisting of a circular core centered on the star, 
plus straight lines emerging from the centroid at angles of approximately 
$45^{\circ}$, $135^{\circ}$, $225^{\circ}$ and $315^{\circ}$ from ecliptic north, 
representing diffraction spikes. Transients within these masks are discarded. 
The four sources masked in this way are Mira, $\tau^{4}$ Eri, $\alpha$ Ceti and
V$^{*}$ EG Cet. 0.07\% of the search area is lost.

\subsubsection{Latents}
\label{sec:latents}
The W1 detector suffers from persistence artifacts dubbed `latents', which
appear as faint, diffuse positive excursions above background. A latent will 
occur at the \textit{detector} position of a sufficiently bright `parent' 
source, in the exposure immediately following imaging of the parent. 
Extremely bright stars can also yield `second latents', artifacts similar to 
standard latents, but occurring \textit{two} exposures after the parent was 
imaged. Because the WISE scan direction varies between coadd epochs, latents do
not always appear at the same sky location, and thus are expected to 
contaminate our list of difference image extractions.

Using the AllWISE catalog and each epochal coadd's unWISE \verb|-frames| 
metadata table\footnote{http://unwise.me/data/-README.txt}, we can predict the 
exact positions of all latents in each coadd. We adopt a latent parent 
brightness threshold of \verb|w1mpro|$\le$8.3, and a second latent threshold of 
\verb|w1mpro|$\le$6.2. We discard any transient within a 1 pixel radius of a 
predicted first or second latent position. $\sim$75,000 persistence artifacts 
are thereby removed from our transient list, while masking only 0.03\% of the 
search area.

W1 latents are much more diffuse than the most streaked Planet
Nine profile for which we claim to be sensitive, with $n_{eff}$ 
$>$10$\times$ that of a point source. Our extraction of tens of thousands of 
latents from the difference images demonstrates our ability to detect even 
highly non-pointlike transients.

\subsubsection{Merging Per-coadd Transient Lists}
\label{sec:resolve}
All of the previous filtering steps are applied on a per-coadd basis. 
However, there will be some transients near, but still within, the edges of 
multiple \verb|coadd_id| footprints, since these are not mutually
exclusive. Because the minimum Moon avoidance maneuver timespan is 0.4 days and
WISE visits are spaced by $\sim$6 months, transients at the same location but 
separated temporally by $<$0.4 days, even if extracted from different coadd 
images, will generally be found in the same set of L1b exposures and are 
therefore not distinct. These multiple detections should be merged prior to 
orbit linking, to avoid the possibility of a transient being linked with 
another copy of itself. We create a single merged transient catalog from the 
per-coadd catalogs by implementing a ``resolve'' step. The resolve step looks 
for groups of transients separated spatially by $\le$1$''$ and temporally by 
$<$0.4 days, retaining only the single transient furthest from the boundary of 
its coadd tile.

\subsubsection{Filtering Summary}
\begin{figure} 
 \begin{center}
  \epsfig{file=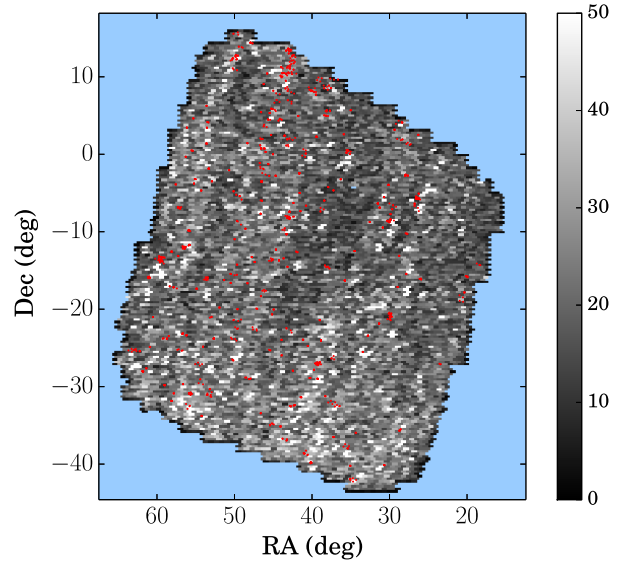, width=3.2in}
  \caption{\label{fig:num_transient_map} Number of transients in the filtered 
catalog per 0.21 square degree $N_{side}$=128 HEALPix pixel. Red dots mark the 
locations of transients linked into low-$\chi^2$ Keplerian quadruplets 
($\S$\ref{sec:ben}).}
 \end{center}
\end{figure}

After filtering, the final transient catalog contains 206,891 detections,
corresponding to a mean density of 113 per square degree. Figure 
\ref{fig:mag_hist} shows a histogram of the transient W1 magnitudes,
which peaks near $W1$=16.6. The number of transients per 
$N_{side}$=128 HEALPix pixel \citep[0.21 square degrees,][]{healpix} is 
shown in Figure \ref{fig:num_transient_map}.

The regions masked due to our various filters are not mutually
exclusive, and in total we have masked 1.0\% of the search area. This
small loss of area is accounted for in the Monte Carlo analysis of 
$\S$\ref{sec:completeness}.

\section{Orbit Linking}
\label{sec:orbfit}
Our basic orbit linking approach is to enumerate all viable quadruplets of 
transients, and determine which can be well fit by a Keplerian orbit. The 
following subsections provide additional details of our orbit linking 
methodology, which is largely inspired by that of \cite{brown15}.

\subsection{Astrometry}
The goal of our astrometric calibration is to ensure that the most 
appropriate and accurate spatial coordinates are input to our orbit linker 
($\S$\ref{sec:orbfit}). The orbit linker computes model orbits in 
ICRS coordinates, and therefore, as described in $\S$\ref{sec:diffem} we have
calibrated the astrometry of each epochal coadd to ICRS via 2MASS. 

It is crucial to have a full accounting of the uncertainties on
our astrometric measurements of transients, since these uncertainties
will strongly influence the $\chi^2$ values of our best-fit orbits.
We include three terms in our astrometric uncertainty model.

\subsubsection{2MASS Calibration Residuals}
We employ the SCAMP outputs \verb|ASTRRMS1|, \verb|ASTRRMS2| as 
per-coadd measures of the residual astrometric scatter between our calibrated 
epochal coadds and 2MASS, for relatively bright sources. The typical value of 
these parameters is 0.28$''$. We aim to derive a \textit{per-coordinate}
astrometric uncertainty, but only one value per transient, rather than
two distinct values for the RA and Dec directions. Therefore, we
define $\sigma_{calib}$=(\verb|ASTRRMS1|+\verb|ASTRRMS2|)/2. We calculate one 
such $\sigma_{calib}$ value per epochal coadd, which is assigned to all 
transients within that coadd. 

\subsubsection{Systematic Proper Motion Shifts}
\label{sec:pm}
We have not corrected for proper motion when calibrating WISE to 2MASS. 
As discussed in \cite{cutri13}, there is a systematic error associated with 
this procedure, because calibrator sources in a given sky region may move
coherently, with nonzero average proper motion\footnote{wise2.ipac.caltech.edu/docs/release/allwise/expsup/sec2\_5.html}.
From \cite{cutri13}, we adopt a maximum systematic error (per coordinate) 
of 200 mas between 2000.0 and 2010.0, and scale this linearly based on the MJD 
of each transient. We call this quantity $\sigma_{pm}$, and its mean value is 
0.26$''$.

\subsubsection{Statistical Scatter}
Statistical scatter due to background noise is by far the dominant
source of astrometric uncertainty for our typical transient. We use the
300,000 injections described in $\S$\ref{sec:completeness} to fit
an analytic relation to the (point source) per-coordinate statistical
scatter as a function of W1 magnitude and integer coverage $n_{exp}$.
We find that the results are well-fit by:

\begin{equation}
\label{equ:scatter}
\sigma_{stat,PSF} = 0.41''\times\frac{\sqrt{12/n_{exp}}}{(10^{(16.0-W1)/2.5})^{0.69}}
\end{equation}

To be conservative, we wish to scale this uncertainty up so that it is 
appropriate even in the case of maximum intra-coadd parallax (8.25$''$).
Typically, for fixed amount of signal, the astrometric uncertainty scales 
linearly with both FWHM and the effective amount of noise. The FWHM 
($n_{eff}$) for 8.25$''$ of parallax is 1.30$\times$ (1.48$\times$) larger than
that of a W1 point source, and therefore we adopt 
$\sigma_{stat}$=1.30$\times$$\sqrt{1.48}$$\times$$\sigma_{stat,PSF}$. Finally, if
the positional error ellipse semi-major axis quoted by Source Extractor is 
larger than our model for $\sigma_{stat}$, we assign $\sigma_{stat}$ to the 
Source Extractor value.

\subsubsection{Total Astrometric Uncertainty}
For our per-coordinate uncertainty provided to the orbit linker 
($\sigma_{ast}$), we adopt the quadrature sum of $\sigma_{calib}$, 
$\sigma_{pm}$ and $\sigma_{stat}$. The median $\sigma_{ast}$ value is
0.90$''$. The $\sigma_{calib}$ and $\sigma_{pm}$ terms effectively provide
a floor at $\sigma_{ast}$$\approx$0.4$''$, even for very bright transients.

\subsection{Transient Linking Rules}
\label{sec:rules}
When enumerating quadruplets, we do not simply allow all transients in the
merged catalog to be paired with all other transients. We enforce 
two rules restricting which transients may be linked:

\begin{enumerate}
\item No two transients may be linked if they are extracted from the same
(\verb|coadd_id|, \verb|epoch|) coadd image.
\item No two transients may be linked if they are separated temporally by 
$<$0.4 days. Because of the way that our coadd epochs are determined from WISE 
visit and Moon avoidance maneuver boundaries, coadd pixels with mean MJD 
separated by less than 0.4 days will in general have been constructed from 
(at least partially) the same set of single-exposure images.
\end{enumerate}

Otherwise all transients may be linked. Importantly, we allow linkages of 
transients which occur during the same season but are extracted from different
coadd images, provided the second rule above is not violated. Note also
that we impose no constraints on the relative fluxes of transients that may be 
linked.

\subsection{Fitting All Quadruplets}
\label{sec:ben}
Our basic orbit linking approach will be to enumerate all viable quadruplets of
transients, and determine which can be well fit by a Keplerian orbit. We break 
this process up into two phases, a``prescreen'' of a large number of possible 
candidates, followed by a detailed fit to Keplerian orbital parameters. The 
prescreen step starts with the complete list of $n_{transient}$ transients, for 
each one identifying a region of the sky where that object might be at 
subsequent epochs. The actual shape of this window is set by our choice of a 
minimum orbital distance, which we take to be 250~AU, and limits on the 
inclination. Our results of $\S$\ref{sec:results} place no restrictions on 
inclination. In this way we create $n_{transient}$ lists of neighboring 
transients, with typically 40 neighbors each. 

The next phase of our prescreen is to run through the neighbor lists, combining
them to form unique quadruplets. We fit the observation times and sky positions
 of each quadruplet with a simple linear model that includes the magnitude of 
the parallax and the 2-D plane-of-sky orbital drift; this model is appropriate 
to distant Solar System bodies observed on time scales that are short compared 
with their orbital periods. If this model yields a $\chi^2 <80$, we accept the 
quadruplet; otherwise we reject it. 

The second phase of our linking algorithm is to fit each prescreened quadruplet
with Keplerian orbital parameters. For this step we use the 
\cite{bernstein00} code, \verb|orbfit|, designed for objects at Kuiper belt
distances and beyond. If \verb|orbfit| successfully resolves a Keplerian 
orbit (see \S\ref{sec:thresh}), we save the quadruplet. Finally we take
the list of saved quadruplets and attempt to combine them into quintuplets, 
sextuplets, etc., if they have transients in common. In this way we start with 
$\sim$$2\times 10^5$ W1 transients, assess over 14~billion quadruplets, and
perform Keplerian orbit fits to the $\sim$12,000 that pass prescreening. The 
final list contains just under 500 successful quadruplet linkages. We always 
retain every tuple with low \verb|orbfit| $\chi^2$, never making cuts on the 
best-fit orbital elements or relative fluxes of constituent transients.

\subsection{Keplerian Orbit $\chi^2$ Threshold}
\label{sec:thresh}
We must choose a goodness-of-fit threshold defining the boundary between
quadruplets that are well fit versus poorly fit by a Keplerian orbit. We do
so by choosing a $\chi^2$ threshold such that the expected false negative
rate is 0.001. Our \verb|orbfit| quadruplet (quintuplet) fits always have 2 or 
3 (4 or 5) degrees of freedom (DOF). To obtain the desired level of false 
negatives, we therefore choose our $\chi^2$ threshold to be 13.8, 16.3, 18.5, 
20.5 for 2, 3, 4, 5 DOF. The impact of our 0.001 failure rate in fitting orbits
can be accounted for in combination with the simulations of 
$\S$\ref{sec:completeness}, but it leads to a negligible shift in our 90\% 
completeness threshold.

\section{Completeness Analysis}
\label{sec:completeness}

\subsection{Notation}
There are two types of completeness relevant to our search. The first 
completeness refers to the probability that a transient source of a given 
brightness is successfully detected in a reference-subtracted epochal coadd.
This will be denoted $f_{det}$. Note that $f_{det}$ will be a function of
coadd integer coverage $n_{exp}$. The second completeness is the probability 
with which an object of a particular brightness is detected as a transient in 
$\ge$4 epochal coadd difference images, given the available number of coadd 
epochs and the number of contributing exposures within each coadd epoch. 
Calculating $f_{\ge4}$ is our ultimate goal, because we require that a Planet 
Nine candidate consist of $\ge$4 transients which are linked along a 
low-$\chi^2$ orbit. Calculating $f_{det}$ as a function of integer 
coverage is one step towards determining $f_{\ge4}$.

\subsection{Parameterizing Completeness Curves}
$f_{det}$ is expected to approach unity for very bright transients, and
should decrease to zero for very faint transients, which will be 
indistinguishable from background noise. The situation is similar for 
$f_{\ge4}$. The rejection of transients due to e.g. our bright star masking 
means that $f_{\ge4}$ will asymptote to a value slightly less than unity for 
bright objects.

For these reasons, we parameterize completeness curves, in both the case of
$f_{det}$ and $f_{\ge4}$, with the model:

\begin{equation}
\label{equ:erf}
f(W1) = \frac{A}{2} \times [1 + \textrm{erf}(-(W1-\mu)/\sigma)]
\end{equation}

\noindent
Where $W1$ is the W1 magnitude, and there are 3 free parameters: $A$, $\mu$
and $\sigma$. $A$ is the asymptotic completeness value toward low $W1$, and is 
typically very close to unity. When fitting ($A$, $\mu$, $\sigma$) we impose 
the constraint $A \le 1$. $\mu$ is the $W1$ value at which the completeness is 
$A/2\approx0.5$, the center of the transition from $f$$\approx$1 to $f$=0. 
$\sigma$ dictates the width of this transition.

\subsection{Computing $f_{det}$ Curves}
\label{sec:f_det}
In order to most accurately calculate $f_{\ge4}$, it is necessary to compile
a set of completeness curves $f_{det}(W1; n_{exp})$, for each value of 
$n_{exp}$. We have done so by injecting 300,000 pointlike fakes of varying $W1$ 
into our difference images. 100 real difference images spread throughout our 
search footprint are used as the basis for 3,000 distinct mock difference 
images, each with 100 fakes injected. We record whether each fake is recovered 
using a 1 pixel match radius. We then bin the fakes by $n_{exp}$, and fit 
Equation \ref{equ:erf} to the fraction of fakes recovered as a function of $W1$ 
for each $n_{exp}$ value.

Not all $n_{exp}$ values are sufficiently sampled by these fakes, because many 
$n_{exp}$ values represent just a tiny fraction of our search area. Most of the 
fakes have $6 \le n_{exp} \le 15$, and these are the $n_{exp}$ values for which 
high-quality $f_{det}$ curves can be created. Table \ref{tab:n_exp} lists the
best-fit completeness curve parameters ($A$, $\sigma$, $\mu$) for these 
$n_{exp}$ values.

\begin{table} [ht]
\centering
\caption{Parameters tabulated as a function of $n_{exp}$, the
number of exposures contributing to a coadd pixel}\label{tab:n_exp}
\begin{tabular}{ c|c|c|c|c|c }
       $n_{exp}$ & $p(n_{exp})$ & $\mu$ & $\sigma$ & $A$ & $\mu_{pred}$\\
 \hline
           3 & 0.0104 & - & - & - & 16.156 \\
           4 & 0.0108 & - & - & - & 16.312 \\
           5 & 0.0117 & - & - & - & 16.433 \\
           6 & 0.0139 & 16.512 & 0.302 & 0.9970 & 16.532 \\
           7 & 0.0197 & 16.624 & 0.356 & 0.9914 & 16.616 \\
           8 & 0.0338 & 16.689 & 0.368 & 0.9940 & 16.688 \\
           9 & 0.0657 & 16.766 & 0.347 & 0.9955 & 16.752 \\
          10 & 0.118 & 16.832 & 0.342 & 0.9940 & 16.809 \\
          11 & 0.175 & 16.867 & 0.352 & 0.9933 & 16.861 \\
          12 & 0.196 & 16.908 & 0.354 & 0.9945 & 16.908 \\
          13 & 0.162 & 16.948 & 0.355 & 0.9943 & 16.952 \\
          14 & 0.0986 & 16.997 & 0.360 & 0.9943 & 16.992 \\
          15 & 0.0450 & 17.044 & 0.368 & 0.9958 & 17.030 \\
          16 & 0.0193 & - & - & - & 17.065 \\
          17 & 9.81$\times$10$^{-3}$ & - & - & - & 17.098 \\
          18 & 4.70$\times$10$^{-3}$ & - & - & - & 17.129 \\
          19 & 2.29$\times$10$^{-3}$ & - & - & - & 17.158 \\
          20 & 1.27$\times$10$^{-3}$ & - & - & - & 17.186 \\
          21 & 7.12$\times$10$^{-4}$ & - & - & - & 17.212 \\
          22 & 5.42$\times$10$^{-4}$ & - & - & - & 17.238 \\
\end{tabular}
\end{table}

One might hope that $\mu$($n_{exp}$) scales as expected from 
statistical noise proportional to $1/\sqrt{n_{exp}}$:

\begin{equation}
\label{equ:mu_pred}
\mu_{pred}(n_{exp}) = \mu(n_{exp,0}) + 2.5\textrm{log}_{10}(\sqrt{n_{exp}/n_{exp,0}})
\end{equation}

We take $n_{exp,0}$=12, given that 12 is both the median
and mode of $n_{exp}$ in our full set of epochal coadds. 
$\mu_{pred}(n_{exp})$ calculated according to
Equation \ref{equ:mu_pred} is also listed in Table \ref{tab:n_exp}, and it 
clearly agrees very well with the measured $\mu(n_{exp})$ values for 
6 $ \le n_{exp} \le$ 15.

For our downstream Monte Carlo analysis, we employ $f_{det}(W1;n_{exp})$ 
completeness curves parameterized by Equation \ref{equ:erf}. For these
parameterizations, we adopt $\mu$=$\mu_{pred}(n_{exp})$. Since $\sigma$
and $A$ appear roughly constant over the range of $n_{exp}$ for which these
have been measured, we adopt their median values ($A$=0.9943, $\sigma$=0.355) 
for all $f_{det}(W1; n_{exp})$ completeness curves during Monte Carlo 
simulations.

\subsection{Coverage Statistics}
Our Monte Carlo simulation of $f_{\ge4}$ requires as input statistics regarding 
the fraction of sky area in the full search footprint as a function of number 
of coadd epochs ($n_{epochs}$) and number of exposures per epochal coadd pixel 
($n_{exp}$). Tables \ref{tab:n_epochs} and \ref{tab:n_exp} list these values,
referred to as $p(n_{epochs})$ and $p(n_{exp})$ respectively. Note that our
analysis discards the small number of coadd pixels with integer coverage 
$\le$2, as it is not possible to perform outlier rejection with fewer than 
three exposures.

\begin{table} [ht]
\centering
\caption{Fraction of pixels in search region as a function of number of
coadd epochs}\label{tab:n_epochs}
\begin{tabular}{ c l|c l }
       $n_{epochs}$ & $p(n_{epochs})$ & $n_{epochs}$ & $p(n_{epochs})$\\
 \hline
       1 & 1.32$\times$10$^{-5}$ & 6 & 0.126 \\
       2 & 1.20$\times$10$^{-5}$ & 7 & 0.483 \\
       3 & 1.41$\times$10$^{-5}$ & 8 & 0.327 \\
       4 & 1.50$\times$10$^{-5}$ & 9 & 0.0635 \\
       5 & 2.50$\times$10$^{-4}$ & & 
\end{tabular}
\end{table}

\subsection{$f_{\ge4}$ Monte Carlo Calculation}
\label{sec:multinomial}
We generate a large number of fake objects, each assigned its own magnitude 
$W1$. Each object is randomly assigned a number of coadded epochs using a 
multinomial draw from $p(n_{epochs})$. Each epoch of each object is then
randomly assigned a corresponding number of exposures $n_{exp}$ using
a multinomial draw from $p(n_{exp})$. Given an object's assigned $W1$
and $n_{exp}$ within each epoch, each appareance of that object will
be detected with a probability given by $f_{det}(W1; n_{exp})$, which
is specified by the parameters ($\mu$, $\sigma$, $A$) described in
the final paragraph of $\S$\ref{sec:f_det}.
Each epoch of each object is then marked as either `detected' or 
`undetected' by performing a Bernoulli trial with success probability given
by $f_{det}(W1; n_{exp})$. $f_{\ge4}(W1)$ is then calculated
by computing the fraction of objects for which $\ge$4 total detections are
accumulated, using a sample of 10,000 objects per value of $W1$, with 
$W1$=16.20,16.21, ..., 17.59, 17.60. 

\begin{figure}
 \begin{center}
  \epsfig{file=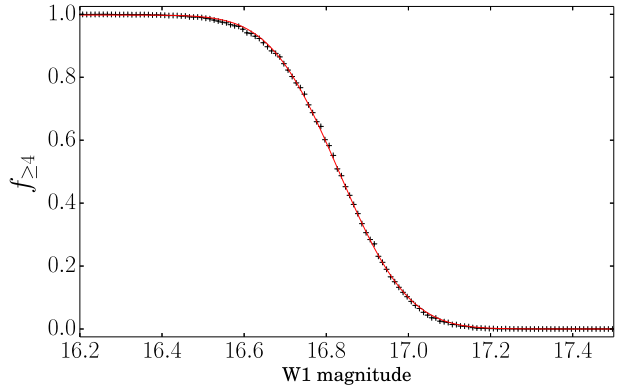, width=3.2in}
  \caption{\label{fig:completeness} Completeness of obtaining $\ge$4 detections
as a function of W1 magnitude. Black plus marks are results of the Monte 
Carlo simulation described in $\S$\ref{sec:multinomial}. Red is the best-fit erf
model, with parameters ($\mu$, $\sigma$, $A$) = (16.83, 0.19, 0.998). 
The completeness curve shown incorporates corrections for area lost to spatial 
masking and non-pointlike morphology.}
 \end{center}
\end{figure}

Our measured $f_{\ge4}$ completeness curve is plotted in Figure 
\ref{fig:completeness}. Fitting the tabulated $f_{\ge4}(W1)$ values with 
Equation \ref{equ:erf} gives ($\mu$, $\sigma$, $A$) = (16.89, 0.18, 0.998). 
This best-fit model reaches $f_{\ge4}$=0.90 (`90\% completeness') at $W1$=16.72.
Requiring five rather than four detections results in a 0.11 magnitude 
reduction in sensitivity, with 90\% completeness at $W1$=16.61. We can account 
for the masking of 1.0\% of our search area ($\S$\ref{sec:filter}) by randomly 
zeroing out 1\% of $f_{det}(W1; n_{exp})$ probabilities during the Monte Carlo. 
We find that this pushes the 90\% completeness threshold brighter by 6.4 (9.5) 
mmag in the case of demanding four (five) linked detections. 

\begin{figure}
 \begin{center}
  \epsfig{file=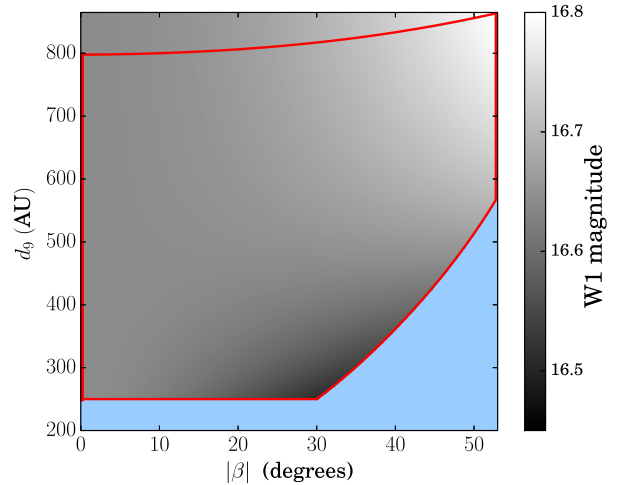, width=3.2in}
  \caption{\label{fig:completeness_map} Variation of our 90\% completeness
threshold for acquiring a quadruplet of detections, due to changes in
amount of parallactic smearing and depth of coverage with $d_9$ and $|\beta|$. 
The red boundary encloses the parameter space accessible to our search, 
assuming the maximally W1-luminous \cite{Fortney16} model. Within the
accessible parameter space, the average 90\% completeness threshold is 16.66 
(16.54) for quadruplets (quintuplets).}
 \end{center}
\end{figure}

We can scale these measured point source completeness thresholds in
a manner consistent with the final term of Equation \ref{equ:depth}, to 
account for variations in sensitivity as a function of $|\beta|$ and $d_9$. The 
resulting 90\% completeness threshold is shown in Figure 
\ref{fig:completeness_map}. Over the parameter space in which we are sensitive 
to the maximally W1-luminous \cite{Fortney16} model, 
the mean 90\% completeness threshold is 16.66 (16.54) when demanding at least 
four (five) linked detections. The $\sim$0.05 mag degradation in sensitivity 
relative to our point source measurement is a result of parallactic smearing, as
the average $\pi_9$ value within our search space is 3.6$''$.

We caution that we might fail to link detections of Planet Nine if it saturates
in W1 ($W1 \lesssim 8.25$), though this would require a luminosity hundreds of 
times greater than any of the \cite{Fortney16} models. We might also
fail to detect Planet Nine if it is brighter than $W1$=9.5, due to the AllWISE
catalog masking described in $\S$\ref{sec:bright}. Finally, we have not 
presented a detailed accounting of small-scale variations in our 90\% 
completeness threshold, such as those introduced by bright star masking and the
spatially varying number of coadd epochs shown in Figure \ref{fig:num_epochs}.

\section{Results}
\label{sec:results}
Orbit linking yields 478 Keplerian quadruplets that pass our $\chi^2$ cut 
described in $\S$\ref{sec:thresh}. The locations of all successfully linked 
quadruplets are indicated by red dots in Figure \ref{fig:num_transient_map}. We
visually inspect finder charts for all 478 candidate quadruplets, plus an 
additional 786 higher-$\chi^2$ quadruplets up to \verb|orbfit| $\chi^2$ of 35.

In situations where all four transients have the same parity, there is 
very nearly zero relative parallax between our coadd epochs. Given this
type of cadence, with observations spaced at integer numbers of years,
Planet Nine's apparent motion could be similar to that of a high proper 
motion star, if close to aphelion with large enough $d_9$ and $e_9$. As a 
result, 113 of our 478 low-$\chi^2$ quadruplets are single-parity linkages 
associated with known high proper motion stars, according to SIMBAD 
\citep{simbad}. These known objects have $|\mu|$ ranging from $\sim$200 mas/yr
to $\sim$2,500 mas/yr. As expected, we do not obtain low-$\chi^2$ cross-parity 
linkages for these fast-moving stars, as their small parallaxes 
($\lesssim$0.1$''$) correspond to completely unbound trajectories of the type 
which \verb|orbfit| is not intended to recover.

All quadruplets not associated with known moving objects appear to be 
affected, at least in part, by one or more problems with the transient catalog.
Among these, the most common contaminants are:

\begin{itemize}
\item Diffraction spikes not fully captured by the masks of 
$\S$\ref{sec:bright}.
\item Kernel matching artifacts that evaded the filtering of 
$\S$\ref{sec:direct}.
\item Residuals of moderately bright static sources just far enough from
the centroid to escape the 2MASS matching of $\S$\ref{sec:tmass}.
\item Extractions affected by the wings of bright stars, falling just
outside of the $\S$\ref{sec:bright} masks.
\item Glints near extremely bright 
stars\footnote{wise2.ipac.caltech.edu/docs/release/allsky/expsup/sec4\_4g.html}.
\item Variable static sources too faint to be captured by our 2MASS 
rejection step.
\end{itemize}

\begin{figure*}
 \begin{center}
  \epsfig{file=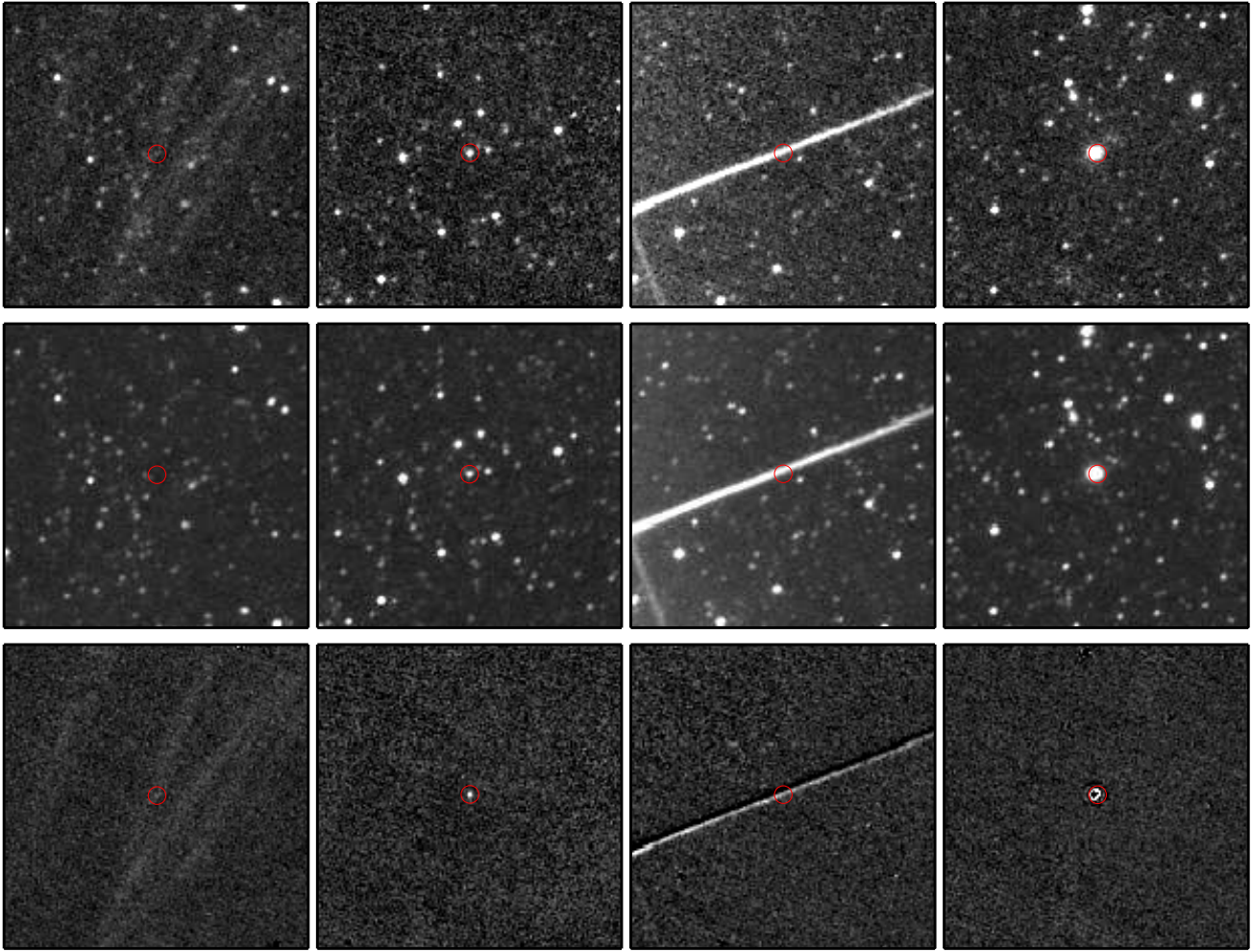, width=7.0in}
  \caption{\label{fig:finder} Examples of common contaminants drawn from the
transients contributing to our 478 low-$\chi^2$ Keplerian quadruplets. Top 
row: epochal coadds. Middle row: corresponding reference images. Bottom row:
HOTPANTS subtractions. Cutouts are 6.9$'$$\times$6.9$'$. Red
circles are centered on the locations of linked difference image
extractions. Left column: nebulosity associated with a glint. 
Middle left column: real variability of a faint, static W1 source not 
cataloged by 2MASS. Middle right column: a diffraction spike not fully masked 
by our prescription of $\S$\ref{sec:bright}. Right column: bright static source 
residuals.}
 \end{center}
\end{figure*}

Examples of several such features are shown in Figure \ref{fig:finder}. As we 
do not find any plausible candidate among linked quadruplets, we quote a 
quadruplet-based limit of $W1$$\ge$16.66 (90\% completeness) over the parameter
space searched. Admittedly, it is not possible to precisely characterize the 
effect of visual inspection on our completeness. 

We obtain 12 low-$\chi^2$ quintuplets. All are single-parity
linkages in which all five transients are associated with a unique
known high proper motion star. Therefore we can rule out the presence of Planet
Nine candidates in the parameter space searched with $W1$$<$16.54 at 90\% 
completeness, independent of any visual inspection.

Our recourse to visual inspection is a deeply regrettable and unsatisfying 
aspect of the present search, and in future extensions of this work we will seek
to eliminate this step. We estimate that by ignoring finders associated with
known fast-moving stars and better masking diffraction spikes, we can reduce
the number of finder chart inspections per unit search area by a factor of 
$\sim$2. Therefore, extending to the $\sim$80\% of sky amenable to our search 
will likely entail inspection of $\sim$4,400 quadruplet finders.

\section{Conclusion}
\label{sec:conclusion}

We have developed a method to search for Planet Nine using custom, inertial
coadds of WISE and NEOWISE-Reactivation W1 exposures. These epochal coadds 
extend much deeper than individual W1 exposures, and remove most of the inner 
solar system objects and instrumental artifacts that would otherwise overwhelm 
such a search. We have demonstrated our method by searching $\sim$2000 square 
degrees of sky identified as a potentially likely present-day Planet Nine 
location. We do not find any plausible candidates, but are able to characterize
our completeness, thereby ruling out the presence of Planet Nine in the 
parameter space searched at $W1$$<$16.66 (90\% completeness).

Translating this value to an optical equivalent is both model and distance 
dependent, because the observed flux from reflected sunlight scales
like $d_9^{-4}$, whereas that from intrinsic emission at 3-4$\mu$m scales like
$d_9^{-2}$. Adopting the maximally W1-luminous \cite{Fortney16} model and
$d_9$=650 AU, $W1$=16.66 corresponds to an optical VR magnitude of 22.06. 
However, alternative models can have far bluer (VR$-$$W1$) colors
at the same distance, and in such cases the corresponding optical constraint 
will be much weaker. For instance, assuming a (VR$-$$W1$) color characteristic 
of the \cite{Fortney16} Table 2 model with $T_{eff}$=50 K, CH$_4$ abundance 
of 10$^{-3}$$\times$ solar and $d_9$=650 AU, the corresponding limit is 
VR$\ge$19.16.

In the near future, we will apply our methodology to all of the W1 sky
area over which we expect substantial sensitivity. We anticipate significantly
reduced sensitivity near the Galactic plane ($|b_{gal}| \lesssim 10^{\circ}$), 
due to the high number density of bright static sources and elevated background 
levels. Also, as illustrated in Figure \ref{fig:mask}, we expect very little
sensitivity at $|\beta|$$\gtrsim$60$^{\circ}$ given our current prescription for
segmenting exposures into coadd epochs. However, by modifying our time-slicing 
rules near the ecliptic poles, we should in principle be able to derive 
deeper than usual constraints in these regions thanks to the 
disproportionately large number of WISE exposures at high $|\beta|$. Our 
methodology should be readily applicable to WISE W2, although the models of 
\cite{Fortney16} suggest that a W2-detectable Planet Nine would likely require 
a relatively large mass (at least $\sim$35$M_{\oplus}$), in tension with 
dynamical constraints \citep{brown16}.

The entire high-probability Planet Nine sky region singled out by \cite{dlfm16}
will be searchable with our method, as that footprint has both low $|\beta|$ and
high $|b_{gal}|$. We will prioritize a search of that region, only 5\% of which
overlaps the footprint considered in this work.

Finally, because our analysis proceeds from WISE single-exposure images, we can
naturally improve upon the present results by generating additional coadd
epochs from future NEOWISER data releases.

\acknowledgments{
We thank Roc Cutri for responding to queries at the WISE Help Desk. We thank
Dustin Lang, Peter Eisenhardt and Mike Brown for their feedback.

This research made use of the NASA Astrophysics Data System (ADS) and the IDL 
Astronomy User's Library at Goddard.\footnote{Available at 
\texttt{http://idlastro.gsfc.nasa.gov}}

This research makes use of data products from the Wide-field Infrared
Survey Explorer, which is a joint project of the University of California, Los
Angeles, and the Jet Propulsion Laboratory/California Institute of Technology,
funded by the National Aeronautics and Space Administration. This research
also makes use of data products from NEOWISE, which is a project of the Jet
Propulsion Laboratory/California Institute of Technology, funded by the
Planetary Science Division of the National Aeronautics and Space
Administration.
This publication makes use of data products from the Two Micron All Sky 
Survey, which is a joint project of the University of Massachusetts and the 
Infrared Processing and Analysis Center/California Institute of Technology, 
funded by the National Aeronautics and Space Administration and the National 
Science Foundation.
}

\bibliographystyle{apj}
\bibliography{p9w1_cetus.bib}

\begin{thebibliography}{31}
\expandafter\ifx\csname natexlab\endcsname\relax\def\natexlab#1{#1}\fi

\bibitem[{{Alard}(2000)}]{alard00}
{Alard}, C. 2000, \aaps, 144, 363

\bibitem[{{Batygin} \& {Brown}(2016)}]{batygin16}
{Batygin}, K., \& {Brown}, M.~E. 2016, \aj, 151, 22

\bibitem[{{Bernstein} \& {Khushalani}(2000)}]{bernstein00}
{Bernstein}, G., \& {Khushalani}, B. 2000, \aj, 120, 3323

\bibitem[{{Bertin}(2006)}]{scamp}
{Bertin}, E. 2006, in Astronomical Society of the Pacific Conference Series,
  Vol. 351, Astronomical Data Analysis Software and Systems XV, ed.
  C.~{Gabriel}, C.~{Arviset}, D.~{Ponz}, \& S.~{Enrique}, 112

\bibitem[{{Bertin} \& {Arnouts}(1996)}]{sextractor}
{Bertin}, E., \& {Arnouts}, S. 1996, \aaps, 117, 393

\bibitem[{{Bertin} {et~al.}(2002){Bertin}, {Mellier}, {Radovich}, {Missonnier},
  {Didelon}, \& {Morin}}]{terapix}
{Bertin}, E., {Mellier}, Y., {Radovich}, M.,  {et~al.} 2002, in Astronomical
  Society of the Pacific Conference Series, Vol. 281, Astronomical Data
  Analysis Software and Systems XI, ed. D.~A. {Bohlender}, D.~{Durand}, \&
  T.~H. {Handley}, 228

\bibitem[{{Brown} {et~al.}(2015){Brown}, {Bannister}, {Schmidt}, {Drake},
  {Djorgovski}, {Graham}, {Mahabal}, {Donalek}, {Larson}, {Christensen},
  {Beshore}, \& {McNaught}}]{brown15}
{Brown}, M.~E., {Bannister}, M.~T., {Schmidt}, B.~P.,  {et~al.} 2015, \aj, 149,
  69

\bibitem[{{Brown} \& {Batygin}(2016)}]{brown16}
{Brown}, M.~E., \& {Batygin}, K. 2016, \apjl, 824, L23

\bibitem[{{Brown} {et~al.}(2004){Brown}, {Trujillo}, \& {Rabinowitz}}]{brown04}
{Brown}, M.~E., {Trujillo}, C., \& {Rabinowitz}, D. 2004, \apj, 617, 645

\bibitem[{{Cowan} {et~al.}(2016){Cowan}, {Holder}, \& {Kaib}}]{Cowan16}
{Cowan}, N.~B., {Holder}, G., \& {Kaib}, N.~A. 2016, \apjl, 822, L2

\bibitem[{{Cutri} {et~al.}(2015){Cutri}, {Mainzer}, {Conrow}, {Masci}, {Bauer},
  {Dailey}, {Kirkpatrick}, {Fajardo-Acosta}, {Gelino}, {Grillmair}, {Wheelock},
  {Yan}, {Harbut}, {Beck}, {Wittman}, {Wright}, {Masiero}, {Grav}, {Sonnett},
  {Nugent}, {Kramer}, {Stevenson}, {Eisenhardt}, {Fabinsky}, {Tholen}, {Papin},
  {Fowler}, \& {McCallon}}]{cutri15}
{Cutri}, R.~M., {Mainzer}, A., {Conrow}, T.,  {et~al.} 2015, {Explanatory
  Supplement to the NEOWISE Data Release Products}, Tech. rep.

\bibitem[{{Cutri} {et~al.}(2011){Cutri}, {Wright}, {Conrow}, {Bauer},
  {Benford}, {Brandenburg}, {Dailey}, {Eisenhardt}, {Evans}, {Fajardo-Acosta},
  {Fowler}, {Gelino}, {Grillmair}, {Harbut}, {Hoffman}, {Jarrett},
  {Kirkpatrick}, {Liu}, {Mainzer}, {Marsh}, {Masci}, {McCallon}, {Padgett},
  {Ressler}, {Royer}, {Skrutskie}, {Stanford}, {Wyatt}, {Tholen}, {Tsai},
  {Wachter}, {Wheelock}, {Yan}, {Alles}, {Beck}, {Grav}, {Masiero}, {McCollum},
  {McGehee}, \& {Wittman}}]{cutri11}
{Cutri}, R.~M., {Wright}, E.~L., {Conrow}, T.,  {et~al.} 2011, {Explanatory
  Supplement to the WISE Preliminary Data Release Products}, Tech. rep.

\bibitem[{{Cutri} {et~al.}(2013){Cutri}, {Wright}, {Conrow}, {Fowler},
  {Eisenhardt}, {Grillmair}, {Kirkpatrick}, {Masci}, {McCallon}, {Wheelock},
  {Fajardo-Acosta}, {Yan}, {Benford}, {Harbut}, {Jarrett}, {Lake}, {Leisawitz},
  {Ressler}, {Stanford}, {Tsai}, {Liu}, {Helou}, {Mainzer}, {Gettings},
  {Gonzalez}, {Hoffman}, {Marsh}, {Padgett}, {Skrutskie}, {Beck}, {Papin}, \&
  {Wittman}}]{cutri13}
---. 2013, {Explanatory Supplement to the AllWISE Data Release Products}, Tech.
  rep.

\bibitem[{{de la Fuente Marcos} \& {de la Fuente Marcos}(2016)}]{dlfm16}
{de la Fuente Marcos}, C., \& {de la Fuente Marcos}, R. 2016, arXiv:1607.05633

\bibitem[{{Fienga} {et~al.}(2016){Fienga}, {Laskar}, {Manche}, \&
  {Gastineau}}]{Fienga16}
{Fienga}, A., {Laskar}, J., {Manche}, H., \& {Gastineau}, M. 2016, \aap, 587,
  L8

\bibitem[{{Fortney} {et~al.}(2016){Fortney}, {Marley}, {Laughlin},
  {Nettelmann}, {Morley}, {Lupu}, {Visscher}, {Jeremic}, {Khadder}, \&
  {Hargrave}}]{Fortney16}
{Fortney}, J.~J., {Marley}, M.~S., {Laughlin}, G.,  {et~al.} 2016, \apjl, 824,
  L25

\bibitem[{{Goldstein} {et~al.}(2015){Goldstein}, {D'Andrea}, {Fischer},
  {Foley}, {Gupta}, {Kessler}, {Kim}, {Nichol}, {Nugent}, {Papadopoulos},
  {Sako}, {Smith}, {Sullivan}, {Thomas}, {Wester}, {Wolf}, {Abdalla},
  {Banerji}, {Benoit-L{\'e}vy}, {Bertin}, {Brooks}, {Carnero Rosell},
  {Castander}, {da Costa}, {Covarrubias}, {DePoy}, {Desai}, {Diehl}, {Doel},
  {Eifler}, {Fausti Neto}, {Finley}, {Flaugher}, {Fosalba}, {Frieman},
  {Gerdes}, {Gruen}, {Gruendl}, {James}, {Kuehn}, {Kuropatkin}, {Lahav}, {Li},
  {Maia}, {Makler}, {March}, {Marshall}, {Martini}, {Merritt}, {Miquel},
  {Nord}, {Ogando}, {Plazas}, {Romer}, {Roodman}, {Sanchez}, {Scarpine},
  {Schubnell}, {Sevilla-Noarbe}, {Smith}, {Soares-Santos}, {Sobreira},
  {Suchyta}, {Swanson}, {Tarle}, {Thaler}, \& {Walker}}]{goldstein15}
{Goldstein}, D.~A., {D'Andrea}, C.~B., {Fischer}, J.~A.,  {et~al.} 2015, \aj,
  150, 82

\bibitem[{{G{\'o}rski} {et~al.}(2005){G{\'o}rski}, {Hivon}, {Banday},
  {Wandelt}, {Hansen}, {Reinecke}, \& {Bartelmann}}]{healpix}
{G{\'o}rski}, K.~M., {Hivon}, E., {Banday}, A.~J.,  {et~al.} 2005, \apj, 622,
  759

\bibitem[{{Holman} \& {Payne}(2016)}]{Holman16}
{Holman}, M.~J., \& {Payne}, M.~J. 2016, arXiv:1604.03180

\bibitem[{{Lang}(2014)}]{lang14}
{Lang}, D. 2014, \aj, 147, 108

\bibitem[{{Linder} \& {Mordasini}(2016)}]{linder16}
{Linder}, E.~F., \& {Mordasini}, C. 2016, \aap, 589, A134

\bibitem[{{Luhman}(2014)}]{luhman14}
{Luhman}, K.~L. 2014, \apj, 781, 4

\bibitem[{{Mainzer} {et~al.}(2014){Mainzer}, {Bauer}, {Cutri}, {Grav},
  {Masiero}, {Beck}, {Clarkson}, {Conrow}, {Dailey}, {Eisenhardt}, {Fabinsky},
  {Fajardo-Acosta}, {Fowler}, {Gelino}, {Grillmair}, {Heinrichsen}, {Kendall},
  {Kirkpatrick}, {Liu}, {Masci}, {McCallon}, {Nugent}, {Papin}, {Rice},
  {Royer}, {Ryan}, {Sevilla}, {Sonnett}, {Stevenson}, {Thompson}, {Wheelock},
  {Wiemer}, {Wittman}, {Wright}, \& {Yan}}]{neowiser}
{Mainzer}, A., {Bauer}, J., {Cutri}, R.~M.,  {et~al.} 2014, \apj, 792, 30

\bibitem[{{Mainzer} {et~al.}(2011){Mainzer}, {Bauer}, {Grav}, {Masiero},
  {Cutri}, {Dailey}, {Eisenhardt}, {McMillan}, {Wright}, {Walker}, {Jedicke},
  {Spahr}, {Tholen}, {Alles}, {Beck}, {Brandenburg}, {Conrow}, {Evans},
  {Fowler}, {Jarrett}, {Marsh}, {Masci}, {McCallon}, {Wheelock}, {Wittman},
  {Wyatt}, {DeBaun}, {Elliott}, {Elsbury}, {Gautier}, {Gomillion}, {Leisawitz},
  {Maleszewski}, {Micheli}, \& {Wilkins}}]{neowise}
{Mainzer}, A., {Bauer}, J., {Grav}, T.,  {et~al.} 2011, \apj, 731, 53

\bibitem[{{Meisner} \& {Finkbeiner}(2014)}]{meisner14}
{Meisner}, A.~M., \& {Finkbeiner}, D.~P. 2014, \apj, 781, 5

\bibitem[{{Meisner} {et~al.}(2016){Meisner}, {Lang}, \& {Schlegel}}]{meisner16}
{Meisner}, A.~M., {Lang}, D., \& {Schlegel}, D.~J. 2016, arXiv:1603.05664

\bibitem[{{Sheppard} \& {Trujillo}(2016)}]{sheppard16}
{Sheppard}, S.~S., \& {Trujillo}, C. 2016, arXiv:1608.08772

\bibitem[{{Skrutskie} {et~al.}(2006){Skrutskie}, {Cutri}, {Stiening},
  {Weinberg}, {Schneider}, {Carpenter}, {Beichman}, {Capps}, {Chester},
  {Elias}, {Huchra}, {Liebert}, {Lonsdale}, {Monet}, {Price}, {Seitzer},
  {Jarrett}, {Kirkpatrick}, {Gizis}, {Howard}, {Evans}, {Fowler}, {Fullmer},
  {Hurt}, {Light}, {Kopan}, {Marsh}, {McCallon}, {Tam}, {Van Dyk}, \&
  {Wheelock}}]{skrutskie06}
{Skrutskie}, M.~F., {Cutri}, R.~M., {Stiening}, R.,  {et~al.} 2006, \aj, 131,
  1163

\bibitem[{{Trujillo} \& {Sheppard}(2014)}]{trujillo14}
{Trujillo}, C.~A., \& {Sheppard}, S.~S. 2014, \nat, 507, 471

\bibitem[{{Wenger} {et~al.}(2000){Wenger}, {Ochsenbein}, {Egret}, {Dubois},
  {Bonnarel}, {Borde}, {Genova}, {Jasniewicz}, {Lalo{\"e}}, {Lesteven}, \&
  {Monier}}]{simbad}
{Wenger}, M., {Ochsenbein}, F., {Egret}, D.,  {et~al.} 2000, \aaps, 143, 9

\bibitem[{{Wright} {et~al.}(2010){Wright}, {Eisenhardt}, {Mainzer}, {Ressler},
  {Cutri}, {Jarrett}, {Kirkpatrick}, {Padgett}, {McMillan}, {Skrutskie},
  {Stanford}, {Cohen}, {Walker}, {Mather}, {Leisawitz}, {Gautier}, {McLean},
  {Benford}, {Lonsdale}, {Blain}, {Mendez}, {Irace}, {Duval}, {Liu}, {Royer},
  {Heinrichsen}, {Howard}, {Shannon}, {Kendall}, {Walsh}, {Larsen}, {Cardon},
  {Schick}, {Schwalm}, {Abid}, {Fabinsky}, {Naes}, \& {Tsai}}]{wright10}
{Wright}, E.~L., {Eisenhardt}, P.~R.~M., {Mainzer}, A.~K.,  {et~al.} 2010, \aj,
  140, 1868

\end{thebibliography}

\end{document}